\documentclass[conference]{IEEEtran}

\usepackage[T1]{fontenc}
\usepackage[utf8]{inputenc}
\usepackage{cite}
\usepackage{amsmath,amssymb}
\usepackage{graphicx}
\usepackage{url}
\usepackage{hyperref}
\usepackage{listings}
\usepackage{xcolor}
\usepackage{booktabs}
\usepackage{microtype}
\usepackage{multirow}
\usepackage{booktabs}
\usepackage{tabularx}
\usepackage{colortbl}

\lstdefinestyle{pythonstyle}{
    language=Python,
    basicstyle=\ttfamily\small,
    keywordstyle=\color{blue},
    stringstyle=\color{magenta!80},
    commentstyle=\color{gray},
    showstringspaces=false,
    numbers=left,
    numberstyle=\small\color{gray},
    stepnumber=1,
    numbersep=6pt,
    frame=single,
    framesep=4pt,
    rulecolor=\color{black!20},
    captionpos=b,
    xleftmargin=5pt,
    xrightmargin=5pt,
    literate={~}{{\textasciitilde}}1,
}
\begin{document}

\title{Typify: A Tool for Usage-driven Type Inference in Python}

\author{
\IEEEauthorblockN{
Hassan Aman~~~Ali Aman~~~Muhammad Asaduzzaman~~~Shaowei Wang$^{\dagger}$~~~Jessica Chen
}

\IEEEauthorblockA{
School of Computer Science, University of Windsor, Canada
}

\IEEEauthorblockA{
$^{\dagger}$Department of Computer Science, University of Manitoba, Canada
}
\IEEEauthorblockA{
\{amanh, burkia, masaduzz, xjchen\}@uwindsor.ca, shaowei.wang@umanitoba.ca
}
}

\maketitle

\begin{abstract}
Python's dynamic type system, while expressive and flexible, creates
significant challenges for static analysis tools in unannotated or
partially annotated codebases. Existing approaches either require
pre-existing type annotations, or depend on deep learning models that
demand large training corpora and significant computational resources.
We present \textsc{Typify}, a lightweight, usage-driven static type
inference engine that infers precise type information across entire
Python projects without annotations or training data. \textsc{Typify}
integrates symbolic execution, fixpoint analysis, and context-matching
retrieval to propagate types across modules, functions, and recursive
call structures. The tool is available as both a command-line interface (CLI) for batch
inference, evaluation and downstream integration with other tools, and a
Visual Studio Code (VS Code) extension for interactive type assistance. Evaluation on the ManyTypes4Py and Typilus
datasets shows that \textsc{Typify} outperforms all static baselines
and closely matches state-of-the-art hybrid deep learning approaches,
while running approximately 10$\times$ faster.

\smallskip\noindent
\textbf{Screencast:} \url{https://youtu.be/ANkhv1dci3s}\\
\textbf{Website:} \url{https://ali-aman.ca/typify}
\end{abstract}

\section{Introduction}
\label{sec:introduction}
Modern software development increasingly relies on Python, yet Python's dynamic type system means that most real-world code carries no type annotations. Empirical studies show that fewer than 10\% of annotatable code elements are typed in practice~\cite{digrazia2022}. Without type information, tools such as linters, static analyzers, and IDEs cannot provide reliable support, and bugs that could be caught early surface only at runtime. While Python's gradual typing standards~\cite{pep484} allow developers to add annotations voluntarily, doing so manually in large or legacy codebases is tedious and error-prone.

Several categories of tools attempt to address this gap. Static type checkers such as Mypy~\cite{mypy2026}, Pyre~\cite{pyre2026}, and Pyright~\cite{pyright2026} are effective when annotations are already present, but fall back to \texttt{Any} when they are not. Deep learning-based tools such as Type4Py~\cite{mir2022type4py} and Typilus~\cite{allamanis2020typilus} achieve broad coverage but are non-deterministic, require large training corpora, and perform poorly on project-specific or user-defined types. Hybrid approaches like HiTyper~\cite{peng2022hityper} combine static reasoning with neural predictions, but operate at the function level and inherit the reproducibility limitations of the models they rely on.

In this paper, we present \textsc{Typify}, a tool for usage-driven static type inference in Python. \textsc{Typify} infers types by reasoning about how variables, functions, and data structures are used across a project, rather than relying on annotations or training data. It builds a project-level dependency graph and propagates type information across modules using fixpoint analysis. For unresolved slots, a context-matching retrieval step fills the remaining gaps without sacrificing determinism. \textsc{Typify} is available as both a command-line interface (CLI) and a Visual Studio Code (VS Code) extension. A complete description of the underlying technique is available in our technical paper~\cite{aman2026typify}.

\section{Approach}
\label{sec:approach}
\textsc{Typify} operates in four stages. We briefly describe each stage below. Full algorithmic details, including formal definitions and threshold selection, can be found elsewhere~\cite{aman2026typify}.

\subsection{Dependency Graph Construction}

Before inference begins, \textsc{Typify} scans the project root and identifies all Python modules, including those nested within packages. Each module is assigned a fully qualified name and represented as a node in a dependency graph, where an edge $M_i \rightarrow M_j$ indicates that $M_i$ imports $M_j$. The graph is built by traversing each module's abstract syntax tree (AST) and extracting its import statements. This gives \textsc{Typify} a precise map of inter-module relationships that governs the order and scope of all downstream inference.

Real-world projects often contain circular imports. \textsc{Typify} detects strongly connected components (SCCs) in the dependency graph and treats each SCC as a unified analysis region, iterating inference within it until all module snapshots stabilize.

\subsection{Scheduling and Fixpoint Resolution}

Once the dependency graph is built, \textsc{Typify} schedules modules for analysis. For projects without circular imports, modules are processed in topological order, so that any types inferred in a dependency are available when its dependents are analyzed. For cyclic dependencies, \textsc{Typify} iterates over each SCC, comparing each module's type snapshot after every pass. Analysis terminates when no snapshot changes, guaranteeing convergence. In practice, most projects complete in a single forward pass.

\subsection{Usage-Driven Inference}

Usage-driven inference is the core of \textsc{Typify}. Rather than relying on annotations or training data, \textsc{Typify} infers types directly from how variables, functions, and data structures are used. All inferred types are represented as \texttt{TypeExpr} instances, a uniform recursive structure capturing a base type and any nested type arguments.

\begin{lstlisting}[language=Python]
class TypeExpr:
    base: TypeExpr
    args: list[TypeExpr]
\end{lstlisting}

For each module, \textsc{Typify} constructs a symbol table and traverses the module body statement by statement. Assignments, function calls, and return statements all contribute evidence that refines the inferred type of each type slot. Type updates are monotonic: types only become more specific or remain unchanged, so analysis always converges. Figure~\ref{lst:example} illustrates this process.

\begin{figure}[h]
\centering
\begin{lstlisting}[language=Python]
def collect_items(flag):
    items = []             # Step 2
    if flag:
        items.append(1)    # Step 3
    else:
        items.append("x")  # Step 4
    return items           # Step 5

collect_items(True)        # Step 1
\end{lstlisting}
\caption{Example of usage-driven and call-site-driven inference.}
\label{lst:example}
\end{figure}

\begin{enumerate}
    \item The call site \texttt{collect\_items(True)} provides concrete evidence that \texttt{flag} is \texttt{bool}, which \textsc{Typify} propagates to the parameter slot before entering the function body.
    \item \texttt{items = []} establishes a provisional type of \texttt{list[Any]}.
    \item \texttt{items.append(1)} refines the slot to \texttt{list[int]}.
    \item \texttt{items.append("x")} unifies the existing type with \texttt{list[str]}, producing \texttt{list[Union[int, str]]}.
    \item The return statement binds the inferred type of \texttt{items} to the function's return slot, giving \texttt{list[Union[int, str]]}.
\end{enumerate}

\subsection{Context-Matching Retrieval}

After usage-driven inference, some type slots may remain unknown, for example in uncalled functions or sparsely used variables. For these cases, \textsc{Typify} applies a context-matching retrieval step. It queries a pre-built index of (context, type) pairs extracted from an annotated codebase, matching each uninferred slot against indexed entries by the similarity of their surrounding source context. Retrieved candidates are treated as suggestions rather than definitive answers, and they never override a type already established by usage-driven inference. \textsc{Typify} assigns three confidence levels to its predictions: high for usage-driven results, mid for retrieval matches, and low for name-convention heuristics applied when retrieval yields no result.

\section{Architecture}
\label{sec:architecture}
\begin{figure}[t]
  \centering
  \makebox[\columnwidth][c]{%
    \includegraphics[width=1.0\columnwidth]{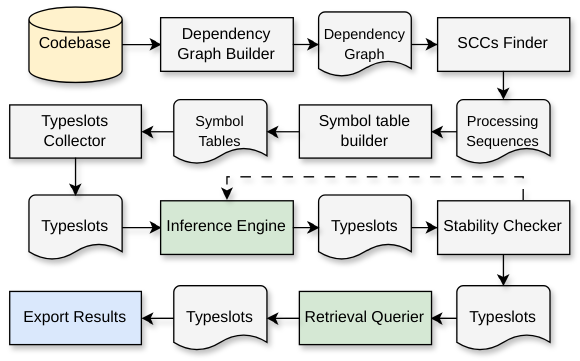}%
  }
  \caption{An overview of \textsc{Typify}'s architecture.}
  \label{fig:architecture}
\end{figure}

Figure~\ref{fig:architecture} shows an overview of \textsc{Typify}'s architecture. The pipeline begins with the Dependency Graph Builder, which scans the project and constructs a module-level import graph, establishing the order and scope of all downstream analysis. The SCCs Finder identifies strongly connected components in this graph to handle circular imports, grouping mutually dependent modules into unified analysis regions. The Symbol Table Builder then constructs a symbol table for each module, mapping identifiers to their scopes and definitions. These are passed to the Typeslots Collector, which extracts all type slots — the specific locations in the code where a type is to be inferred. The Inference Engine performs usage-driven type inference over these slots, propagating types from call sites, assignments, and return statements. The Stability Checker determines whether the fixpoint has been reached by comparing type snapshots across passes, feeding updated typeslots back into the Inference Engine if not. Once stable, the Retrieval Querier fills any remaining uninferred slots by querying a pre-built index of annotated code for contextually similar entries. Finally, the Export Results component writes the inferred types to JSON artifacts consumed by downstream tools or the VS Code extension.

\section{Implementation}
\label{sec:implementation}
\textsc{Typify} is available in two forms: a command-line interface (CLI) for batch inference and evaluation, and a Visual Studio Code (VS Code) extension for interactive use.

\begin{table}[t]
\caption{Summary of \textsc{Typify} CLI Commands. More details of these commands can be found in~\cite{typify-cli}.}
\label{tab:cli}
\centering
\small
\begin{tabular}{@{}p{\columnwidth}@{}}
\toprule

\texttt{\textbf{typify infer} <project-path> <output-path> [{-}{-}config <config-path>]} \\[2pt]
Runs inference on a project and writes inferred types to the output file. \\[3pt]
\midrule

\texttt{\textbf{typify gt} <dataset-path> <output-path>} \\[2pt]
Extracts ground-truth type annotations from an annotated dataset. \\[3pt]
\midrule

\texttt{\textbf{typify dataset} <dataset-path> <output-path> [{-}{-}config <config-path>]} \\[2pt]
Runs batch inference over a dataset and saves predictions for evaluation. \\[3pt]
\midrule

\texttt{\textbf{typify eval} <gt-path> <tool-path>} \\[2pt]
Evaluates predictions against ground truth using type-matching metrics. \\[3pt]
\midrule

\texttt{\textbf{typify build} <dataset-path> <index-path> [{-}{-}workers <n>]} \\[2pt]
Builds a retrieval index from an annotated Python corpus. \\[3pt]

\bottomrule
\end{tabular}
\end{table}

The \textsc{Typify} inference engine implements the four-stage pipeline described in Section~\ref{sec:approach}. It is distributed as a standalone Python package with a command-line interface, and requires Python 3.11 or later. Installation and invocation are covered in Section~\ref{sec:usage}. The available commands are summarized in Table~\ref{tab:cli}. A \texttt{config.json} file can be passed to the relevant commands via the \texttt{---config} flag, and allows users to adjust parameters of the inference pipeline, including the number of top-$k$ results returned by context-matching retrieval, whether context-matching retrieval is enabled, whether the deep learning model integration is enabled, and other inference parameters.

Given a project directory and an output path, the engine runs the full inference pipeline and writes the inferred types for all type slots to the output directory as JSON artifacts, ready to be consumed by downstream tools.

\textsc{Typify} has the capability of integrating with a deep learning model such as Type4Py~\cite{mir2022type4py}, which serves as a fallback for low-confidence slots. This combination outperforms HiTyper + Type4Py, setting a new state of the art. Context-matching retrieval is powered by Tantivy~\cite{tantivy}, a full-text search engine library. Type4Py queries are handled via the Requests~\cite{requests} HTTP library.

The VS Code extension communicates with the inference engine by invoking the CLI as a subprocess and consuming its JSON output. Inferred types are displayed on hover: hovering over any identifier shows its inferred type, and for function definitions, the hover panel also lists known call sites and their locations. Where a type annotation is missing, the hover panel includes an \texttt{Annotate} button that inserts the inferred type directly into the source file. A sidebar panel exposes the same configuration toggles as \texttt{config.json}; changes write back to the file and trigger re-analysis. When a file changes, the extension re-analyzes only the affected files instead of rerunning inference over the whole project, which keeps the feedback loop fast for continuous use during development.

\section{Usage Scenarios}
\label{sec:usage}
\subsection{Case I: Batch Type Inference via the CLI}

Consider a developer who wants to run type inference over a Python project to feed inferred types into a downstream tool, such as a static analyzer or documentation generator. They install the inference engine via \texttt{pip install typify-cli} and invoke the \texttt{typify infer} command (see Table~\ref{tab:cli} for the full list of available commands):

\begin{lstlisting}
typify infer <project-path> <output-path>
\end{lstlisting}

\textsc{Typify} runs the inference pipeline and writes the inferred types as JSON artifacts. Each entry records the inferred type, its confidence level, and the source location of the type slot. The developer can adjust inference behaviour by editing \texttt{config.json}, toggling context-matching retrieval and Type4Py integration, or adjusting the usage-driven inference parameters to balance accuracy and performance.

\subsection{Case II: Interactive Type Assistance via the VS Code Extension}

\begin{figure*}[t]
  \centering
  \makebox[\columnwidth][c]{%
    \includegraphics[width=2.15\columnwidth]{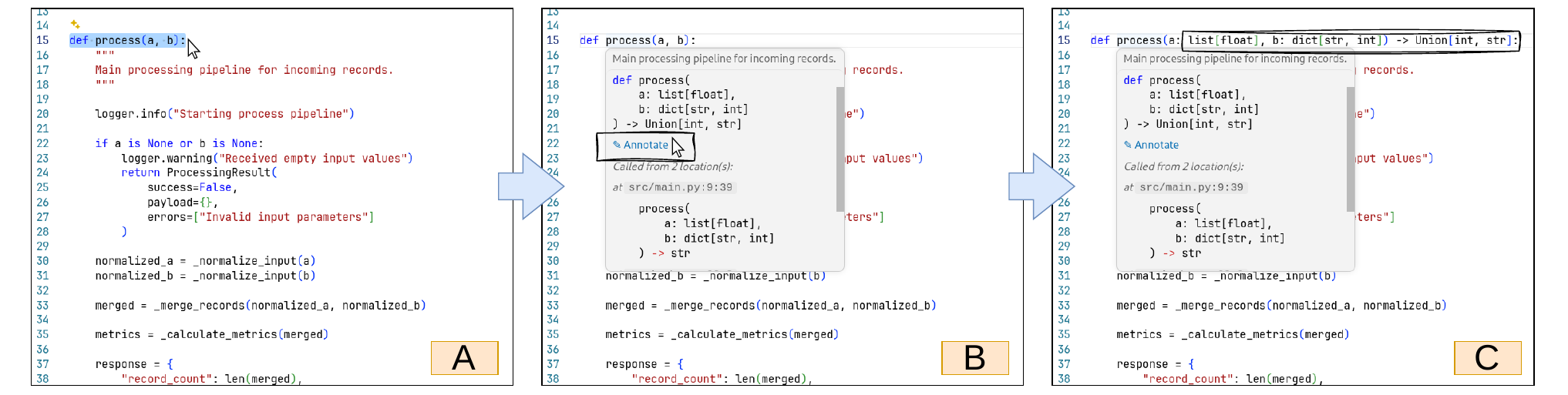}%
  }
  \caption{Using \textsc{Typify} in VS Code: (a) unannotated function in the editor, (b) hover overlay showing inferred types, call sites, and the \textit{Annotate} button, (c) the function after clicking \textit{Annotate} with type annotations inserted.}
  \label{fig:vscode}
\end{figure*}

Consider a developer working on a legacy Python codebase. They encounter a function with no type annotations and want to understand its expected inputs and return type. The developer hovers over the function definition, and the \textsc{Typify} overlay appears, showing inferred types for each parameter and the return value. The panel also lists the known call sites from which these types were derived, including the observed argument types and return value, allowing the developer to verify the evidence behind each inferred type. The developer clicks \textit{Annotate}, and \textsc{Typify} inserts type annotations for the function's parameters and return value directly into the source file. The function is now typed, allowing type mismatches to be caught statically rather than at runtime. Figure~\ref{fig:vscode} illustrates this workflow across all three stages.

\section{Evaluation}
\label{sec:evaluation}
This section summarizes the quantitative evaluation of \textsc{Typify}. Full details of the evaluation procedure and results can be found in the technical paper~\cite{aman2026typify}.

\textbf{Setup.} We evaluate \textsc{Typify} on two datasets: ManyTypes4Py~\cite{mir2021manytypes4py}, a benchmark of Python projects with type annotations, and the Typilus dataset~\cite{allamanis2020typilus}, consisting of annotations collected from 600 Python repositories. We report two metrics: exact match, where a prediction is correct only if it matches the ground truth in both base type and type arguments, and base match, where type arguments are ignored. We compare \textsc{Typify} against three compared techniques: Pyre Infer~\cite{pyre2026}, an industry static type checker developed by Meta; HiTyper~\cite{peng2022hityper}, a hybrid approach that combines static inference with deep learning; and Type4Py~\cite{mir2022type4py}, a deep learning-based tool. We also evaluate \textsc{Typify} + Type4Py, where Type4Py serves as a fallback for low-confidence slots, and compare it against HiTyper + Type4Py.

\textbf{Results.} \textsc{Typify} outperforms all compared techniques by substantial margins on both datasets (Table~\ref{tab:results}). On ManyTypes4Py, it improves over HiTyper's static component by 11.7--16.5 points overall, with especially large gains on argument types (35--40 points), and by 45--56 points over Pyre Infer. \textsc{Typify} also beats Type4Py overall despite being fully static. Combined with Type4Py as a fallback, \textsc{Typify} + Type4Py surpasses HiTyper + Type4Py on most tasks, e.g., 75.0\% vs.\ 74.2\% exact match for variable types, and a 14.7-point gain on argument prediction. The same trend holds on the Typilus dataset, confirming the hybrid combination sets a new state of the art across both benchmarks.

\begin{table}[t]
  \caption{Top-1 type prediction accuracy (\%). EM = exact match, BM = base match.}
  \label{tab:results}
  \centering
  \resizebox{\columnwidth}{!}{%
  \begin{tabular}{llrrrr}
    \toprule
    & & \multicolumn{2}{c}{ManyTypes4Py} & \multicolumn{2}{c}{Typilus} \\
    \cmidrule(lr){3-4} \cmidrule(lr){5-6}
    Task & Approach & EM & BM & EM & BM \\
    \midrule
    \multirow{5}{*}{Variable}
      & \textsc{Typify}             & 62.7 & 74.7 & 18.5 & 55.3 \\
      & \textsc{Typify} + Type4Py   & 75.0 & 86.5 & 19.0 & 65.2 \\
      & HiTyper                     & 57.0 & 63.8 & 16.3 & 52.7 \\
      & HiTyper + Type4Py           & 74.2 & 81.6 & 18.3 & 63.3 \\
      & Type4Py                     & 55.2 & 60.1 & 14.4 & 33.6 \\
      & Pyre Infer                  &  6.9 &  7.1 &  0.9 &  1.1 \\
    \midrule
    \multirow{5}{*}{Argument}
      & \textsc{Typify}             & 43.0 & 47.7 & 49.2 & 52.2 \\
      & \textsc{Typify} + Type4Py   & 60.0 & 65.5 & 65.4 & 69.4 \\
      & HiTyper                     &  7.8 &  8.1 &  7.0 &  7.1 \\
      & HiTyper + Type4Py           & 45.3 & 47.7 & 52.1 & 54.0 \\
      & Type4Py                     & 46.0 & 47.9 & 50.3 & 51.7 \\
      & Pyre Infer                  &  9.4 &  9.4 &  8.8 &  8.8 \\
    \midrule
    \multirow{5}{*}{Return}
      & \textsc{Typify}             & 38.4 & 50.5 & 31.6 & 44.0 \\
      & \textsc{Typify} + Type4Py   & 48.0 & 60.4 & 40.6 & 55.2 \\
      & HiTyper                     & 34.0 & 43.7 & 26.9 & 37.5 \\
      & HiTyper + Type4Py           & 46.8 & 56.9 & 40.7 & 52.9 \\
      & Type4Py                     & 33.9 & 36.4 & 29.4 & 32.4 \\
      & Pyre Infer                  & 35.2 & 36.2 & 27.3 & 28.3 \\
    \bottomrule
  \end{tabular}}
\end{table}

\textbf{Inference Speed.} \textsc{Typify} requires an average of 4.7\,ms per data point on ManyTypes4Py, compared to 48.2\,ms for HiTyper and 36.5\,ms for Type4Py, corresponding to a 90\% reduction in latency relative to HiTyper. A similar trend holds on the Typilus dataset, where \textsc{Typify} requires 3.9\,ms per data point compared to 45.3\,ms for HiTyper. Since \textsc{Typify} is fully static, these figures exclude training time, which Type4Py and HiTyper both require.

\section{User Study}
\label{sec:userstudy}
We conducted a user study to assess the usefulness of \textsc{Typify} and to identify any usability issues. All five participants are Computer Science students with at least three years of Python programming experience. As this is a small-scale study, our results should be read as preliminary evidence rather than a general claim of effectiveness. To avoid the confound of participants struggling with an unfamiliar codebase, we developed a controlled Python project implementing a library management system, covering operations such as checking books in and out, managing loans, and tracking members; we view this toy project as a useful first step, and plan to evaluate \textsc{Typify} on larger, real-world codebases with more participants in future work.

Each participant completed a set of tasks with and without the \textsc{Typify} VS Code extension enabled. Tasks included adding new functionality, adjusting existing functions, and identifying expected types in unfamiliar parts of the code. When working without \textsc{Typify}, participants used Pylance (which uses Pyright), the standard Python type inference environment that ship with the VS Code Python extension. Participants completed tasks without \textsc{Typify} first, then repeated equivalent tasks with the extension enabled, followed by an interview.

All five participants agreed that \textsc{Typify} helped them understand the codebase more quickly. Pylance and Pyright fell back to \texttt{Any} for unannotated functions, giving participants little actionable information. \textsc{Typify} inferred concrete types and surfaced the call sites from which those types were derived, which participants found particularly useful for understanding not just what type was inferred, but why. The \textit{Annotate} button also reduced the friction of adding annotations manually. Without \textsc{Typify}, participants took noticeably longer to reason about types and in several cases introduced type mismatches that would only have surfaced at runtime.

The study also surfaced several areas for improvement. Two participants noted that \textsc{Typify} could not provide linter features such as auto-completion. One participant suggested the sidebar show the Type4Py API endpoint and allow changing it for custom self-hosted instances — we found this valuable and incorporated it. The remaining features are currently under active development.

\section{Related Work}
\label{sec:related}
A number of static type checkers are available for Python, including but not limited to Mypy~\cite{mypy2026}, Pyre~\cite{pyre2026}, Pyright~\cite{pyright2026}, and Pytype~\cite{pytype2026}. However, these tools focus on type checking rather than inference, and often default to \texttt{Any} when annotations are missing, limiting their usefulness for legacy or unannotated codebases.

To address this, several tools infer types using neural models trained on large code corpora. Typilus~\cite{allamanis2020typilus} uses graph neural networks, Type4Py~\cite{mir2022type4py} applies deep similarity learning, DeepTyper~\cite{hellendoorn2018} uses sequence models, and TypeWriter~\cite{pradel2020typewriter} uses search-based validation. More recent work has explored LLM-based annotation generation~\cite{zhang2024generating, peng2023typegen}, and HiTyper~\cite{peng2022hityper} combines static analysis with deep learning over type dependency graphs. These LLM-based assistants generate annotations by predicting plausible tokens from patterns learned during training, rather than by verifying how a value is actually constructed and used, so their output can vary across runs and is not guaranteed to be consistent with the surrounding project. \textsc{Typify} instead derives each annotation deterministically by tracing concrete usage through the project's dependency graph, so a given codebase always yields the same, traceable result. However, existing approaches are either non-deterministic, limited to function-level inference, or struggle with project-specific types. \textsc{Typify} requires no training data, produces deterministic results, and performs whole-project inference, with \textsc{Typify} + Type4Py outperforming HiTyper + Type4Py on most tasks. Beyond Python, gradual typing systems such as TypeScript~\cite{typescript2026} and Flow~\cite{flow2026} combine inference with constraint solving.

\section{Limitations}
\label{sec:limitations}
\textsc{Typify} handles the majority of real-world Python code well, as reflected in its evaluation results across diverse open-source repositories. However, several language features present challenges for purely static analysis. Metaprogramming constructs and decorators that dynamically modify definitions at runtime are not fully supported, and dynamic imports cannot be resolved through the dependency graph. For example, consider the following code:

\begin{lstlisting}
module_name = "os.path"
mod = importlib.import_module(module_name)
result = mod.join("usr", "bin")
\end{lstlisting}

Since the module name is determined at runtime, \textsc{Typify} cannot resolve the type of \texttt{mod} or \texttt{result} statically, and falls back to \texttt{Any} for those slots. Higher-order functions are supported in common cases, but complex patterns involving mutual recursion or context-dependent signatures may not be fully resolved, causing \textsc{Typify} to fall back to \texttt{Any} for the affected slots. Support for several of these features is currently under active development. In practice, these limitations rarely affect usability since the majority of Python codebases do not rely heavily on such constructs.

\section{Conclusion}
\label{sec:conclusion}
We present \textsc{Typify}, a tool for usage-driven static type inference in Python. \textsc{Typify} infers types by reasoning about and tracing how variables, functions, and data structures are used throughout an entire project, without relying on existing type annotations or training data. It is available both as a Visual Studio Code extension for interactive use and as a command-line interface for batch inference and integration with downstream tools. \textsc{Typify} not only outperforms traditional static analysis and deep learning--based approaches in type inference accuracy, but also performs inference significantly faster (e.g., \textsc{Typify} is approximately 10$\times$ faster than HiTyper), making it practical for real-time use in developer tools. The source code and datasets are publicly available, enabling other researchers to replicate and extend our work~\cite{typify-cli}. The demonstration will showcase how \textsc{Typify} can be used in a variety of usage scenarios.

\bibliographystyle{IEEEtran}
\bibliography{refs}

\end{document}